\documentclass[aip,graphicx]{revtex4-1}

\usepackage{graphicx, color}
\usepackage{dcolumn}
\usepackage{bm}
\usepackage{textcomp}
\usepackage{here}
\usepackage{multirow}
\usepackage{amsmath,amssymb,bm}

\draft 

\begin{document}


\title{A Simple and Effective Solution to the Constrained QM/MM Simulations}  



\author{Hideaki Takahashi* }
\email[Hideaki Takahashi: ]{hideaki@m.tohoku.ac.jp}
\affiliation{Department of Chemistry, \\ Graduate School of Science, \\ Tohoku University, \\ Sendai, Miyagi 980-8578, 
Japan}
\author{Hiroyuki Kambe}
\affiliation{Department of Chemistry, \\ Graduate School of Science, \\ Tohoku University, \\ Sendai, Miyagi 980-8578, 
Japan}
\author{Akihiro Morita}
\affiliation{Department of Chemistry, \\ Graduate School of Science, \\ Tohoku University, \\ Sendai, Miyagi 980-8578, Japan}
\affiliation{Elements Strategy Initiative for Catalysts and Batteries (ESICB), \\ Kyoto University,
Kyoto 615-8520, Japan}


\date{\today}

\begin{abstract}

It is a promising extension of the quantum mechanical/molecular mechanical (QM/MM) approach to incorporate the solvent molecules surrounding the QM solute into the QM region to ensure the adequate description of the electronic polarization of the solute. However, the solvent molecules in the QM region inevitably diffuse into the MM bulk during the QM/MM simulation. In this article  we developed a simple and efficient method, referred to as \textquoteleft boundary constraint with correction (BCC)\textquoteright, to prevent the diffusion of the solvent water molecules by means of a constraint potential. The point of the BCC method is to compensate the error in a statistical property due to the bias potential by adding a correction term obtained through a set of QM/MM simulations. The BCC method is designed so that the effect of the bias potential completely vanishes when the QM solvent is identical with the MM solvent. Furthermore, the desirable conditions, that is, the continuities of energy and force and the conservations of energy and  momentum, are fulfilled in principle. We applied the QM/MM-BCC method to a hydronium ion($\rm H_3O^+$) in aqueous solution to construct the radial distribution function(RDF) of the solvent around the solute. It was demonstrated that the correction term fairly compensated the error and led the RDF in good agreement with the result given by an \textit{ab initio} molecular dynamics simulation.

\end{abstract}

\pacs{}

\maketitle 


\section{INTRODUCTION}
The hybrid quantum mechanical / molecular mechanical (QM/MM) approach has been extensively utilized to study  solvation processes in chemistry and physics.\cite{rf:warshel1976jmb, rf:gao1992sci, rf:gao1996acr, rf:ruizlopez2003jmst, Canuto2008, rf:senn2009acid, AQC2010, Rivail2015, rf:takahashi2001jcc, rf:takahashi2003jcp} Actually, it offers a versatile theoretical framework for describing the electronic state of a QM object under the influence of a surrounding MM environment. The polarization of the electron density of the QM region due to the environment is a matter of a critical importance in the QM/MM simulations. The success of the QM/MM method in condensed phase simulations is attributed mainly to the fact that the polarization of the QM solute can be reasonably realized by the Coulomb potential due to simple point charges placed on the MM molecules. We note, however, that the charge-transfer type polarization between QM and MM objects cannot be considered in the QM/MM method at least in a theoretically proper manner. In particular in a hydration of an ion, where the charge-transfer type polarizations play a role, the QM/MM approach is no longer adequate to evaluate the solute-solvent interaction in general. Furthermore the orbital mixings due to the intermolecular charge migration will seriously affect the exchange repulsion\cite{rf:takahashi2015jcp, rf:takahashi2016jcp} between  solute and solvent molecules defined usually by the size parameters in the Lennard-Jones potentials\cite{rf:allen_tildesley}. It is, thus, desirable to develop a methodology to resolve the problems associated with the QM/MM boundary to extend the frontier of the QM/MM approach. 

\indent A straightforward but promising solution to the problems is to incorporate the solvent molecules surrounding the solute of interest into the QM region. This treatment offers an adequate description of the electronic states of the solute interacting with the neighboring solvent molecules. However, such a preferred solvent configuration will collapse during a simulation due to the free diffusion of the QM solvent into the MM bulk solvent.  To prevent the diffusion of the QM solvent, there have been a lot of developments that are classified into two categories referred to as \textquoteleft adaptive\textquoteright\ \cite{Kerdcharoen2002, Rode2006, Heyden2007, Zhang2008, Bulo2009, Bernstein2012, Takenaka2012, Watanabe2014} and \textquoteleft constrained\textquoteright\ \cite{Rowley2012, Shiga2013} QM/MM approaches. In the former approaches, the solvent molecules can change its character adaptively from QM to MM (and vice versa) when they cross the QM/MM boundary. Most of the adaptive approaches introduce a transition(buffer) zone between QM and MM regions to ensure the continuity of the energy and forces of the solvent. In the latter, on the other hand, some constraint forces are applied to the solvent to keep the solvent molecules within a solvation shell.  {\color{black} We note, however, that any constraint or adaptive procedure violates a principle of the statistical mechanics since the situation that some particular solvent molecules stay within a shell is obviously against the entropy effect.} \textcolor{black}{In other words, the mixing entropy will be inevitably lost in a constraint approach.} Thus, the requirement we can impose at most on a boundary treatment is that the artifact completely vanishes when the force field of the QM solvent is identical with that of the MM solvent. This may constitute a principal condition to be fulfilled in the development of the adaptive and the constrained QM/MM methods. Actually, the adaptive QM/MM approach will satisfy this condition since it is naturally expected that the effect of an adaptive procedure will disappear  when the QM solvent is identical to the MM solvent. Rowley and Roux developed the FIRES(Flexible Inner Region Ensemble Separator) method\cite{Rowley2012}, categorized into the constrained QM/MM, which offers a surprisingly simple solution though an infinitely steep constraint should be applied for the requirement to be fulfilled.  A sophistication was, then, made by Shiga and Masia in the approach called BEST(Boundary based on Exchange Symmetry Theory)\cite{Shiga2013} which applies bias force to every pair of the QM and MM particles to prevent the two regions from being mixed. In practice, however, the bias forces will be applied only to the pairs with significant contributions to reduce the large computational costs.  

\indent In the present work we develop a new method of constrained QM/MM, called Boundary Constraint with Correction (BCC), which fulfills the desired conditions. That is, the continuities of energy and force, and the conservations of energy and momentum are guaranteed in the BCC method and more importantly BCC is designed so that the effect of an applied bias potential completely disappears when QM solvent is identical to MM solvent. Furthermore the BCC method can be readily implemented provided some QM/MM code is available. As will be described in the Theory section the basic idea underlying the BCC approach is very simple. The point of the method is to make a correction compensating the error in a statistical property of interest due to the constraint force by performing a set of separate QM/MM simulations. Thus, the BCC method necessitates an additional  procedure in contrast to the FIRES\cite{Rowley2012} and BEST\cite{Shiga2013} methods. However, the computational load associated with the correction is rather modest. The efficiency of our method is first assessed by conducting full MM simulations for water solutions where the water solvent is partitioned into two regions through a bias potential. Then, the radial distribution functions (RDF)\cite{rf:allen_tildesley} obtained by BCC are compared with those yielded in conventional simulations without the bias potential. The robustness of the method is also examined by varying the number of water molecules contained in each domain. We next utilize the BCC method to study the hydration of a hydronium ion in water solution where $\rm H_3O^+$ and water molecules within the first solvation shell are represented by a density functional theory(DFT) for electrons. The RDF for water solvent around $\rm H_3O^+$ is constructed through a QM/MM simulation combined with BCC.   

\indent This article is organized as follows. In the next section we show the theoretical framework of the BCC approach that can be fully expressed with a single equation.  In Section III the computational details for the full MM and the QM/MM simulations are presented. In Section IV the RDFs given by the MM simulations combined with BCC are compared with those obtained by a conventional approach to discuss the efficiency and the robustness of the BCC method. The hydration structure for $\rm H_3O^+$ constructed using the QM/MM$-$BCC method is also compared with a recent result provided by a first-principles simulation. Finally, in Section V we summarize our work and make a remark on the prospect of the extension of BCC to free energy calculations.  

\section{THEORY AND METHODOLOGY}
This section consists of two subsections. In Subsection A the details for the boundary constraint with correction (BCC) will be described. Then, the explicit form for the constraint potential employed in the present work will be given in Subsection B. 
\subsection{Boundary constraint with correction}
\begin{figure}[h]
\scalebox{0.6}[0.6] {\includegraphics[trim=170 150 150 120,clip]{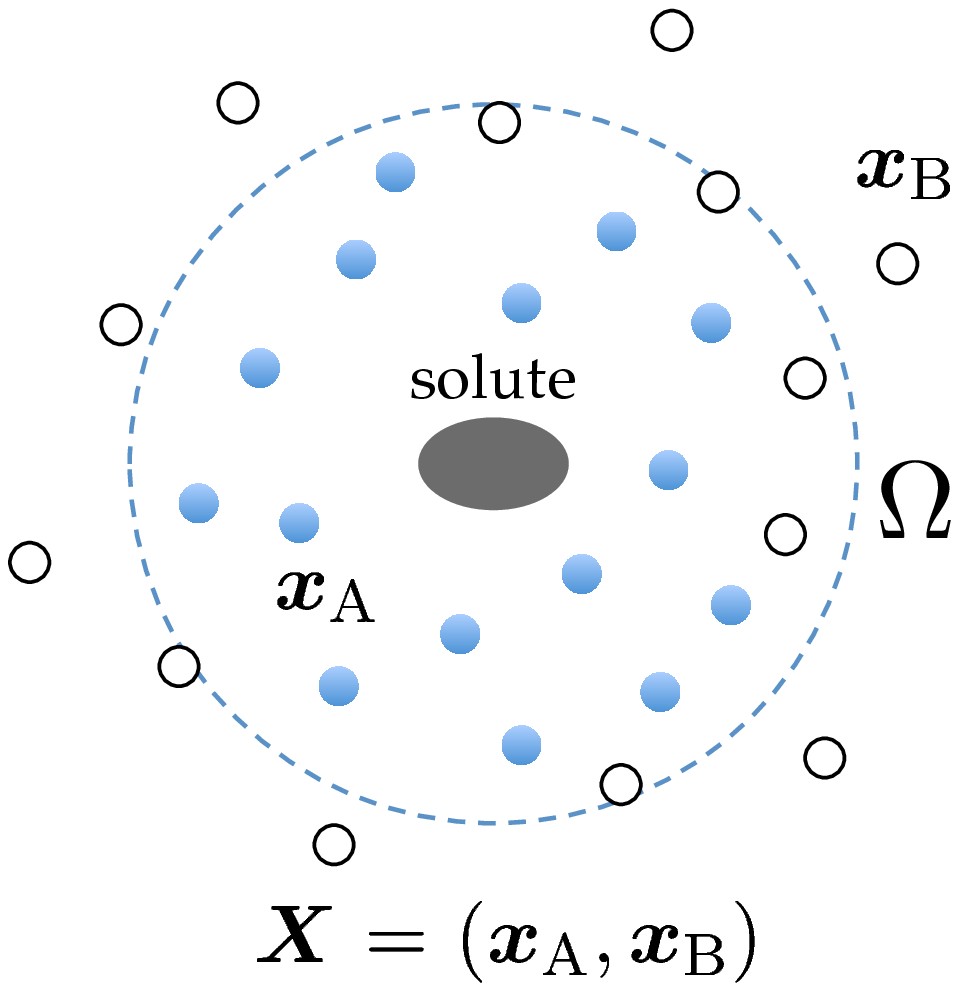}}            
\caption{\label{boundary} Schematized illustration for a molecular simulation with a boundary constraint. The blue and white circles represent  solvent \textsf{A} and \textsf{B}, respectively.  $\bm{x}_{\rm A}$ and $\bm{x}_{\rm B}$ collectively denote the coordinates of the constituent molecules of the solvent \textsf{A} and \textsf{B}, respectively. $\bm{X}$ represents the full set of the coordinates, i.e. $\bm{X}=(\bm{x}_{\rm A},\bm{x}_{\rm B})$. The blue broken circle $\Omega$ shows the solvation shell within which solvent \textsf{A} is confined by a constraint potential. The molecules \textsf{B} are allowed to enter the inner region of the shell. }
\end{figure}

To introduce the basic idea of the BCC approach we consider here a solution described with pairwise potentials. The discussion can also be extended to a QM/MM system straightforwardly. We suppose that the solvent consists of two kinds of molecules \textsf{A} and \textsf{B} with different solute-solvent and solvent-solvent potentials. As illustrated in Fig. \ref{boundary} solvent \textsf{A} is being confined within a solvation shell $\Omega$ enclosing a solute. Then, the number $N_{\rm A}$ of molecules \textsf{A} would be much smaller than $N_{\rm B}$ for the bulk solvent \textsf{B} in general. 
For our present purpose it can be assumed that the force field of molecule \textsf{A} is reasonably close to that of \textsf{B}. Provided that $\bm{x_{\text{A}}}$ collectively denotes the configuration of the solvent molecules \textsf{A} the interaction energy $U_{\rm A}(\bm{x_{\text{A}}})$ can be given by
\begin{equation}
U_{{\rm A}}\left(\bm{x}_{{\rm A}}\right)=\sum_{i}^{N_{{\rm A}}}u_{{\rm A}}\left(\bm{x}_{{\rm A}i}\right)+\sum_{i<j}^{N_{{\rm A}}}\upsilon_{{\rm A}}\left(\bm{x}_{{\rm A}i},\bm{x}_{{\rm A}j}\right)
\label{eq:E_A}
\end{equation}
where $u_{\rm A}$ and $\upsilon_{\rm A}$ are, respectively, the solute-solvent and solvent-solvent interaction potentials for solvent \textsf{A}, and $\bm{x}_{{\rm A}i}$ denotes the full coordinates of $i$th molecule.   In Eq. (\ref{eq:E_A}) it is assumed that the position and the orientation of the solute molecule is being fixed during a simulation. We also have an equivalent equation for solvent \textsf{B}, thus, 
\begin{equation}
U_{{\rm B}}\left(\bm{x}_{{\rm B}}\right)=\sum_{i}^{N_{{\rm B}}}u_{{\rm B}}\left(\bm{x}_{{\rm B}i}\right)+\sum_{i<j}^{N_{{\rm B}}}\upsilon_{{\rm B}}\left(\bm{x}_{{\rm B}i},\bm{x}_{{\rm B}j}\right)   \;\;\; . 
\label{eq:E_B}
\end{equation}   
The interaction $U_{{\rm AB}}\left(\bm{x}_{{\rm A}},\bm{x}_{{\rm B}}\right)$ between solvent \textsf{A} and \textsf{B} is written as 
\begin{equation}
U_{{\rm AB}}\left(\bm{x}_{{\rm A}},\bm{x}_{{\rm B}}\right)=\sum_{i}^{N_{{\rm A}}}\sum_{j}^{N_{{\rm B}}}\upsilon_{{\rm AB}}\left(\bm{x}_{{\rm A}i},\bm{x}_{{\rm B}j}\right)   \;\;\;  . 
\label{eq:E_AB}
\end{equation}
As illustrated in Fig. 1 we apply a bias potential $U_{\rm bias}$ to solvent molecules \textsf{A} to keep them staying within a solvation shell $\Omega$. In terms of these potential energies, the statistical average $\left\langle P\right\rangle _{{\rm bias}}^{u_{{\rm A}},\upsilon_{{\rm A}};u_{{\rm B}},\upsilon_{{\rm B}};\upsilon_{{\rm AB}}}$ of a physical property $P$ under the constraint potential can be expressed by
\begin{equation}
\left\langle P\right\rangle _{{\rm bias}}^{u_{{\rm A}},\upsilon_{{\rm A}};u_{{\rm B}},\upsilon_{{\rm B}};\upsilon_{{\rm AB}}}=\frac{\int d\bm{X}P\left(\bm{X}\right)\exp\left[-\beta\left(U_{{\rm A}}+U_{{\rm B}}+U_{{\rm AB}}+U_{{\rm bias}}\left(\bm{x}_{{\rm A}}\right)\right)\right]}{\int d\bm{X}\exp\left[-\beta\left(U_{{\rm A}}+U_{{\rm B}}+U_{{\rm AB}}+U_{{\rm bias}}\left(\bm{x}_{{\rm A}}\right)\right)\right]} \;\;\;  .
\label{eq:P}
\end{equation}
In Eq. (\ref{eq:P}) $\beta$ is the inverse of the Boltzmann constant $k_{\rm B}$ multiplied by temperature $T$, and the coordinate $\bm{X}$ denotes the set of the coordinates $\bm{x}_{{\rm A}}$ and $\bm{x}_{{\rm B}}$, i.e. $\bm{X} =(\bm{x}_{\rm A}, \bm{x}_{\rm B}) $. We are interested in how to eliminate the contribution due to the bias potential from the statistical quantity $\left\langle P\right\rangle _{{\rm bias}}^{u_{{\rm A}},\upsilon_{{\rm A}};u_{{\rm B}},\upsilon_{{\rm B}};\upsilon_{{\rm AB}}}$. To this end we consider the solution with \textquoteleft homogeneous\textquoteright \; solvent in which solvent \textsf{A} also obeys the potentials $(u_{{\rm B}}, \upsilon_{{\rm B}})$ in Eq. (\ref{eq:E_B}), and $\upsilon_{\rm AB}$ in Eq. (\ref{eq:E_AB}) is identical to $\upsilon_{\rm B}$. Then, the ensemble average $\left\langle P\right\rangle _{{\rm bias}}^{u_{{\rm B}},\upsilon_{{\rm B}}}$ under the bias potential is written as 
\begin{equation}
\left\langle P\right\rangle _{{\rm bias}}^{u_{{\rm B}},\upsilon_{{\rm B}}}=\frac{\int d\bm{X}P\left(\bm{X}\right)\exp\left[-\beta\left(U_{{\rm B}}\left[\bm{X}\right]+U_{{\rm bias}}\left(\bm{x}_{{\rm A}}\right)\right)\right]}{\int d\bm{X}\exp\left[-\beta\left(U_{{\rm B}}\left[\bm{X}\right]+U_{{\rm bias}}\left(\bm{x}_{{\rm A}}\right)\right)\right]}    \;\;\; .
\label{eq:P2}
\end{equation}
For the homogeneous solvent system, the average $\left\langle P\right\rangle ^{u_{{\rm B}},\upsilon_{{\rm B}}}$ without the bias potential is also well defined and merely given by
\begin{equation}
\left\langle P\right\rangle _{{\rm }}^{u_{{\rm B}},\upsilon_{{\rm B}}}=\frac{\int d\bm{X}P\left(\bm{X}\right)\exp\left[-\beta U_{{\rm B}}\left(\bm{X}\right)\right]}{\int d\bm{X}\exp\left[-\beta U_{{\rm B}}\left(\bm{X}\right)\right]}   \;\;\;  .
\label{eq:P3}
\end{equation}
The quantity of Eq. (\ref{eq:P3}) subtracted by Eq. (\ref{eq:P2}) can be regarded as the correction to the physical property $\left\langle P\right\rangle$ to compensate the effect due to the boundary constraint exerted on molecules \textsf{A}. We, thus, define the correction term $\Delta_{{\rm corr}}\left\langle P\right\rangle ^{u_{{\rm B}},\upsilon_{{\rm B}}}$ as
\begin{equation}
\Delta_{{\rm corr}}\left\langle P\right\rangle ^{u_{{\rm B}},\upsilon_{{\rm B}}}=\left\langle P\right\rangle _{{\rm }}^{u_{{\rm B}},\upsilon_{{\rm B}}}-\left\langle P\right\rangle _{{\rm bias}}^{u_{{\rm B}},\upsilon_{{\rm B}}}    \;\;\;  .
\label{eq:corr}
\end{equation}
Therefore, Eq. (\ref{eq:P}) will be corrected as  
\begin{equation}
\left\langle P\right\rangle _{{\rm }}^{u_{{\rm A}},\upsilon_{{\rm A}};u_{{\rm B}},\upsilon_{{\rm B}};\upsilon_{{\rm AB}}}=\left\langle P\right\rangle _{{\rm bias}}^{u_{{\rm A}},\upsilon_{{\rm A}};u_{{\rm B}},\upsilon_{{\rm B}};\upsilon_{{\rm AB}}}+\Delta_{{\rm corr}}\left\langle P\right\rangle ^{u_{{\rm B}},\upsilon_{{\rm B}}}  \;\;\; .
\label{eq:P_corr}
\end{equation}
It is readily recognized in Eq. (\ref{eq:P_corr}) that the effect of the bias potential completely disappears when $u_{\rm A}=u_{\rm B}$ and $\upsilon_{\rm A}=\upsilon_{\rm B}=\upsilon_{\rm AB}$. Equation (\ref{eq:P_corr}) describes the  framework of our boundary constraint approach. As a major application of our method we consider a QM/MM system where the solvent \textsf{A} as well as a solute molecule are described with a quantum chemical theory and the solvent \textsf{B} is expressed with an MM force field. In such a system, Eq. (\ref{eq:P_corr}) can be rewritten as 
\begin{equation}
\left\langle P\right\rangle _{{\rm }}^{E_{{\rm QM}};E_{{\rm MM}};E_{{\rm QM/MM}}}=\left\langle P\right\rangle _{{\rm bias}}^{E_{{\rm QM}};E_{{\rm MM}};E_{{\rm QM/MM}}}+\Delta_{{\rm corr}}\left\langle P\right\rangle ^{E_{{\rm MM}}}    \;\;\;  .
\label{eq:P_QM/MM}
\end{equation}
Of course, in Eq. (\ref{eq:P_QM/MM}), $E_{\rm QM}$ and  $E_{\rm MM}$ are the energies of the QM and MM regions in the system, respectively, and $E_{\rm QM/MM}$ is the interaction between the two regions. 

\indent In closing this subsection we make a remark on the BCC approach. Since the correction in BCC is yielded through the  statistical averaging as shown in Eq. (\ref{eq:corr}) it is not possible to apply it to evaluate some dynamical properties. We also note that the solvent molecules \textsf{B} are allowed to enter the solvation shell since no constraint forces are directly applied to them. However, the solvation shell $\Omega$ will be exclusively occupied by the solvent \textsf{A} and the effect of the contamination due to solvent \textsf{B} will be marginal. Anyway, we should determine the number $N_{\rm A}$ beforehand for the given size of $\Omega$. The effect of the choice of $N_{\rm A}$ will be examined in Results and Discussion. As shown in Eq. (\ref{eq:P_corr}) the BCC energy consists of three ensemble averages, each of which utilizes a trajectory yielded in a continuous molecular dynamics simulation conserving energy and momentum.   

\subsection{Potential for constraint}
In this Subsection we introduce the bias potential $U_{\rm bias}$ employed in the present work to constrain the solvent \textsf{A} within a solvation shell $\Omega$. Actually, the choice of $U_{\rm bias}$ is somewhat arbitrary as long as it well behaves. Our choice in this work is to adopt the logarithm of the Fermi function that was proposed in Ref. \onlinecite{Shiga2013}. The explicit form of $U_{\rm bias}$ is given by 
\begin{equation}
U_{{\rm bias}}\left(r\right)=-k_{B}T\; \text{\ensuremath{\log\frac{1}{1+\exp\left(-\alpha\left(r_{c}-r\right)\right)}}}
\label{eq:bias}
\end{equation}
where $r$ is the solute-solvent distance and $r_c$ is the radius of the spherical shell $\Omega$ assuming that the center of a spherical shell $\Omega$ is placed at an interaction site of the solute. It can be easily recognized that $U_{\rm bias}(r)$ is almost zero in the region of $0\leq r\leq r_{c}$, while the slope of $U_{\rm bias}(r)$ asymptotically approaches $k_B T \times \alpha$ for $r_c \leq r$. Thus, a solvent molecule feels no constraint force when it is inside the shell $\Omega$, while it will be repelled by a slope when it  
exits the shell. A half harmonic potential can also be used as a bias potential $U_{\rm bias}$ though not tested in the present work. 

\indent As described above the shape of the shell $\Omega$ is assumed to be spherical in the present benchmark calculations. In a practical application, however,  a more complex shape will be appropriate depending on the structure of the solute molecule of interest. An instant solution to this problem is for example to make a union of the spherical shells $\{\Omega_{i}\} \left(i=1,2,\cdots\right)$, each of which center is placed at the $i$th interaction site of the solute.  Of course, such a non-spherical shell can be readily adopted in the BCC method.  \textcolor{black}{We note a multiple-sphere scheme was also proposed in the adaptive QM/MM method.\cite{Heyden2007}}     

\section{COMPUTATIONAL DETAILS}
In the following we apply the BCC method to two kinds of aqueous solutions as benchmark calculations. We first consider a solution in which a water molecule is solvated by water solvent. All of the molecules in the system are represented with classical force fields.\cite{rf:allen_tildesley} The solvent consists of two kinds of molecules \textsf{A} and \textsf{B} specified with different potentials. A constrained potential is applied to solvent molecules \textsf{A} to keep them within a solvation shell $\Omega$ around the solute. Then, the radial distribution function (RDF)\cite{rf:allen_tildesley} for the solvent will be obtained through Eq. (\ref{eq:P_corr}). The details of the simulations are provided in Subsection A. Next, the hydration of a hydronium ion($\rm{H}_3\rm{O}^+$) is studied by a QM/MM simulation combined with the BCC method, where a hydronium ion and the neighboring 4 water molecules are involved in the QM region through a bias potential. We also compute the RDF for water molecules around the solute to make comparisons with a reference result given by an \textit{ab initio} molecular dynamics (AIMD) simulation.\cite{Tse2015} The computational details for the QM/MM-BCC simulations are given in Subsection B.    

\subsection{A water molecule embedded in a water solvent}
We apply the BCC method to solutions described with fully classical force fields. In each solution system, a water solute is embedded in a mixed solvent comprised of two kinds of water molecules \textsf{A} and \textsf{B} with different interaction potentials. In applying the BCC method, solvent \textsf{A} is confined within a shell $\Omega$ through a constraint force whereas no bias potential is applied to solvent \textsf{B} as illustrated in Fig. (\ref{boundary}). For all the water molecules in the system the force field parameters except for the point charges refer to the SPC/E model.\cite{rf:berendsen1987jpc} Solvent \textsf{B} is completely identical to SPC/E model and has a negative charge $q_{\rm o}^{\rm B} = -0.8476$ in the unit of the elementary charge on the oxygen site, while in solvent \textsf{A} it is shifted to $q_{\rm o}^{\rm A} = -0.90$ assuming a larger polarization. The solute molecule is assumed to be identical to solvent \textsf{A} and hence the charge $q_{\rm o}$ on the oxygen site is taken as $-0.90$. Of course, the charge on a hydrogen atom is determined to ensure the neutrality of the water molecule. To assess the efficiency of the BCC approach we construct the radial distribution functions(RDF) through Eq. (\ref{eq:P_corr}) for the solvent oxygen around the solute oxygen. The Newtonian equations of motion for the solvent molecules are numerically solved using the leap-frog algorithm\cite{rf:allen_tildesley} and the internal structure of the solvent molecule is fixed during the simulation using the quaternion.\cite{rf:allen_tildesley} The time step for the molecular dynamics(MD) is set at 1 fs. The solute water molecule is fixed at the center of a cubic simulation box with a periodic boundary condition.\cite{rf:allen_tildesley} Total number of the water molecules contained in the simulation box is 500 including the solute. The long-range intramolecular Coulomb interaction is evaluated using the Ewald method.\cite{rf:ewald1921ap, rf:allen_tildesley} The thermodynamic condition is set at temperature $T = 300$ K and density $\rho = 1.0\; \rm {g/cm}^3$. The statistical average for each RDF is yielded from a trajectory of 2 ns MD simulation.      

\indent We conduct the BCC simulations employing the bias potential defined by Eq. (\ref{eq:bias}), where the O$-$O distance between solute and solvent is taken as the variable $r$. Thus, the bias potential $U_{\rm bias}$ is applied only to the oxygen atoms of the solvent. Two values of $r_c$ ($= 3.5$ \AA \; and $7.5$ \AA) are tested to examine the effects of the size of the shell $\Omega$ on the resultant solvent structure. The value $\alpha$ in Eq. (\ref{eq:bias}), which specifies the slope of the bias potential, is set at $\alpha = 10.0\;{\rm a.u.}^{-1}$.  The average number $N_{\rm av}$ of the solvent molecules within the shell $\Omega$ is taken as $N_{\rm A}$ for the corresponding shell. The value of $N_{\rm av}$ is evaluated through a simulation for the solution consisting of pure SPC/E water molecules. Then, $N_{\rm A}$ are obtained as $5$ and $58$ for the shells $\Omega$ with $r_c = 3.5$ \AA\;  and 7.5 \AA, respectively. The total number $N$ of the solvent molecules is fixed at $N=499$ in every simulation. 

\subsection{QM/MM-BCC simulation for hydronium ion}
Next we combine the QM/MM simulation with the BCC method(QM/MM-BCC) to perform a benchmark test for a more practical system. We consider here the solvation of a hydronium ion into an aqueous solution. In our QM/MM-BCC system the QM region consists of $\rm{H}_3\rm{O}^+$ and 4 water molelcules described with the Kohn-Sham density functional theory (KS-DFT),\cite{rf:kohn1965pr, rf:parr_yang_eng} while the MM solvent is represented with SPC/E water molecules.\cite{rf:berendsen1987jpc} The Lennard-Jones parameters\cite{rf:allen_tildesley} for SPC/E is assigned to the oxygen atom of the hydronium ion. The RDF for the solvent oxygen around the solute oxygen is constructed to make comparisons with that given by the first-principles approach.\cite{Tse2015} The molecular geometry of the solute is optimized by the KS-DFT with the B3LYP exchange correlation functional\cite{rf:becke1988pra, rf:becke1993Ajcp, rf:lee1988prb} and the aug-cc-pVTZ basis set.\cite{rf:dunning1989jcp} The optimization is performed by conducting Gaussian 09 program package.\cite{rf:gaussian09} Thus, the geometrical parameters  for $\rm{H}_3\rm{O}^+$ are determined as $R(\rm{OH}) = 0.980$ \AA\;  and $\angle{\rm HOH} = 112.8^{\circ}$.  The position and the geometry of the QM solute are fixed during the QM/MM simulation.   

\indent The QM/MM simulation combined with the BCC procedure is performed using our code \textquoteleft Vmol\textquoteright.\cite{rf:takahashi2000cl, rf:talahashi2001jpca, rf:takahashi2001jcc, rf:takahashi2004jcp} A notable feature of the code is that the electronic state of the QM part of the system is determined by the KS-DFT utilizing the real-space grids\cite{rf:chelikowsky1994prl, rf:chelikowsky1994prb, rf:jing1994prb} to represent the one-electron wave functions as well as the operators in the electronic Hamiltonian. The kinetic energy operator is represented with the fourth-order finite difference approach. The interaction between valence electrons and nuclei is evaluated utilizing the pseudopotentials in the Kleinman and Bylander separable form.\cite{rf:kleinman1982prl} The exchange-correlation energy of the QM system is evaluated with the BLYP functional.\cite{rf:becke1988pra, rf:lee1988prb} The wave functions are contained within a cubic QM cell which has 80 grids along each axis. The grid spacing $h$ is set at 0.166 \AA, which leads the QM cell size $l=13.3$ \AA. The efficient double grid technique developed by Ono and Hirose\cite{rf:ono1999prl} is employed to realize the rapid behaviors of the pseudopotentials near the atomic cores.  The width of the double grid is set at $h/7$. The geometry of the QM water molecule is the same as that of the SPC/E model. \textcolor{black}{The internal coordinates of the QM water molecules as well as the QM solute are being fixed during the simulation.} In the QM/MM-MD simulation the position and the orientation of the QM solvent are updated in the same way as for the MM solvent in the previous subsection III. A. Only the difference is that the forces acting on the QM nuclei are directly determined from the Hellmann-Feynman forces\cite{rf:parr_yang_eng, rf:jing1994prb} described in terms of the electronic wave functions. The parameters $\alpha$ and $r_c$ in Eq. (\ref{eq:bias}) are, respectively, set at $\alpha = 10.0\;{\rm a.u.}^{-1}$ and $r_c=3.5$ \AA. 

\indent The solvent molecules in the MM region consists of 495 water molecules represented with the SPC/E model. The computational details for the molecular dynamics of the QM/MM-BCC simulation refer to those for the simulations in the previous subsection. The RDF for the solvent oxygen around the hydronium oxygen is obtained through Eq. (\ref{eq:P_QM/MM}).  Each ensemble average in Eq. (\ref{eq:P_QM/MM}) is constructed from a trajectory yielded through a 200 ps MD simulation. The thermodynamic conditions are set at $T = 300$ K and $\rho = 1.0\; \rm {g/cm}^3$.     

\section{RESULTS AND DISCUSSIONS}
This section is also partitioned into two subsections in parallel to the organization of the Section \textquoteleft Computational Details\textquoteright. Subsection A is devoted to show the results of the BCC simulations for a water molecule immersed in a water solvent, and in Subsection B the results and discussions are provided for the QM/MM-BCC simulations applied to a hydronium ion in water solution.   

\subsection{A water molecule embedded in a water solvent}
\begin{figure}[h]
\scalebox{1.05}[1.05] {\includegraphics[trim=90 550 70 110,clip]{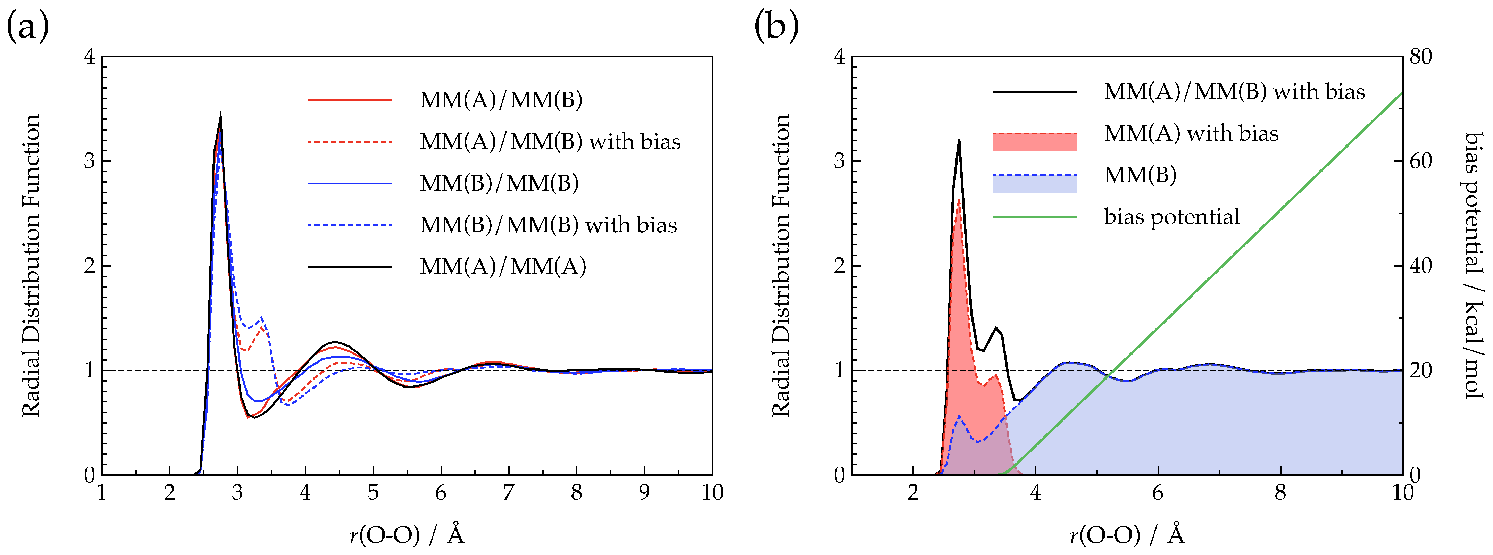}}            
\caption{\label{Rc=35A} (a) The radial distribution functions (RDFs) relevant to the BCC simulation are presented for the oxygen atoms of solvent($\rm{H}_2\rm{O}$) around the solute($\rm{H}_2\rm{O}$) oxygen. The radius of the shell $\Omega$ is set at $r_c=3.5$ \AA. The notations MM(A) and MM(B), respectively, stand for solvent \textsf{A} and \textsf{B} described with molecular mechanical(MM) force fields. The legend \textquoteleft MM(A)/MM(B)\textquoteright\; refers to the RDF constructed by the BCC method. (b) Individual contributions from solvents \textsf{A} and \textsf{B} to the RDF of \textquoteleft MM(A)/MM(B) with bias\textquoteright\; are shown. The bias potential of Eq. (\ref{eq:bias}) is also presented ($\alpha = 10.0\;{\rm a.u.}^{-1}$,  $r_c=3.5$ \AA).}
\end{figure}

As mentioned in Computational Details, all the water molecules in the system refer to the SPC/E model of water except for the values of point charges. The negative charges on the oxygen atoms for the solute and solvent \textsf{A} were set at $-0.90$ in the unit of the elementary charge. For solvent \textsf{B} the original value of $-0.8476$ for SPC/E was adopted to the oxygen atoms in the water molecules. We first apply the bias potential of Eq. (\ref{eq:bias}) with $r_c = 3.5\;$\AA\; to solvent \textsf{A} ($N_{\rm A} = 5$) in the BCC simulations.  Figure \ref{Rc=35A}(a) shows the RDFs needed to evaluate Eq. (\ref{eq:P_corr}) for the solvent oxygen around the solute oxygen.  The legend \textquoteleft ${\rm MM(A)}/{\rm MM(B)}$ with bias\textquoteright\; refers to the RDF obtained with the ensemble average of Eq. (\ref{eq:P}). Similarly \textquoteleft ${\rm MM(B)}/{\rm MM(B)}$ with bias\textquoteright\; and \textquoteleft ${\rm MM(B)}/{\rm MM(B)}$\textquoteright\; are, respectively, the RDFs given by the averages of Eqs. (\ref{eq:P2}) and (\ref{eq:P3}). In the BCC method, these RDFs serve to construct the corrected RDF, \textquoteleft ${\rm MM(A)}/{\rm MM(B)}$\textquoteright\; as described explicitly in Eqs. (\ref{eq:corr}) and (\ref{eq:P_corr}). A notable feature in the RDFs for the solvents under the influence of the bias potential is that unphysical kinks appear due to the potential rising at $r_c = 3.5\;$\AA. However, in the RDF obtained through the BCC procedure, such an unfavorable behavior is fairly eliminated as depicted in the graph of \textquoteleft ${\rm MM(A)}/{\rm MM(B)}$\textquoteright\; by virtue of the correction term of Eq. (\ref{eq:corr}). We now compare the RDF given by BCC with that for the pure SPC/E solvent (\textquoteleft ${\rm MM(B)}/{\rm MM(B)}$\textquoteright). The height of the first peak for the BCC approach is scarcely enhanced as compared with the SPC/E model. However, it is exhibited in the graph that the depth of the depression after the first peak and the height of the second peak are both increased. It is quite natural that the structure of RDF is more emphasized in \textquoteleft ${\rm MM(A)}/{\rm MM(B)}$\textquoteright \; than that for the SPC/E solvent since  water molecule \textsf{A} is being more polarized on purpose than solvent \textsf{B}. Ideally, for our purpose, it is expected that the RDF yielded by the BCC method is coincident with that for the solution with pure solvent \textsf{A} at least in the region within the shell $\Omega$. The RDF for solvent \textsf{A} (\textquoteleft ${\rm MM(A)}/{\rm MM(A)}$\textquoteright), also shown in the Fig. \ref{Rc=35A},  reasonably agrees with the result given by BCC. Thus, it was demonstrated that the RDF deformed by a bias potential can be reasonably refined with the BCC correction. 

\indent In our constraint method solvent \textsf{B} is allowed to enter the inside of the shell $\Omega$. Hence, the ratio of the occupancy of the shell by solvent \textsf{B} is also of our concern. To this end we computed the individual RDFs for solvents \textsf{A} and \textsf{B}. The results are depicted in Fig. \ref{Rc=35A}(b) which also shows the bias potential of Eq. (\ref{eq:bias}) as a function of the O$-$O distance $r$. It is seen in the figure that the inner region of the shell is partly occupied by solvent \textsf{B}. Explicitly, the ratio of the occupancy of solvent \textsf{A} to \textsf{B} is estimated as $\sim 5$ at the first peak of the RDF. Thus, it was revealed that solvent \textsf{B} also enters the inner region of the shell although solvent \textsf{A} dominates the region. This feature constitutes a difference of our approach from other constraint methods such as FIRES\cite{Rowley2012} and BEST\cite{Shiga2013}. 
\indent \textcolor{black}{It is of fundamental interest how the bias potential affects the orientational structure of the solvent molecules around the solute water. To quantify the tetrahedral order we introduce the orientational order parameter  $q$ defined in Ref. \onlinecite{Errington2001}, thus, 
\begin{equation}
q=1-\frac{3}{8}\sum_{i=1}^{3}\sum_{j=i+1}^{4}(\text{\ensuremath{\cos}}\theta_{ij}+\frac{1}{3})^{2} \;\;\; .
\label{eq:OOP}
\end{equation}
$\theta_{ij}$ in Eq. (\ref{eq:OOP}) represents the angle between the lines O$_{\rm x}-$O$_i$ and O$_{\rm x}-$O$_j$ where O$_{\rm x}$ is the oxygen of the solute, while O$_{i}$ and O$_{j}$ are those of the solute's nearest neighbours. The parameter $q$ of Eq. (\ref{eq:OOP}) is designed so that it becomes $1.0$ when a perfect tetrahedral structure is formed around the solute while the mean value of $q$ vanishes at the ideal gas limit. For the systems of  \textquoteleft MM(B)/MM(B)\textquoteright\ and \textquoteleft MM(B)/MM(B) with bias\textquoteright\ $q$ were evaluated as $0.67$ and $0.63$, respectively. We, thus, found that the tetrahedral network is somewhat deformed by the bias potential. The order parameter $q$ for the simulation of \textquoteleft MM(A)/MM(B) with bias\textquoteright\ was obtained as $0.63$ which can be corrected as $0.67$ through the BCC scheme.  }

\begin{figure}[h]
\scalebox{1.05}[1.05] {\includegraphics[trim=90 550 70 100,clip]{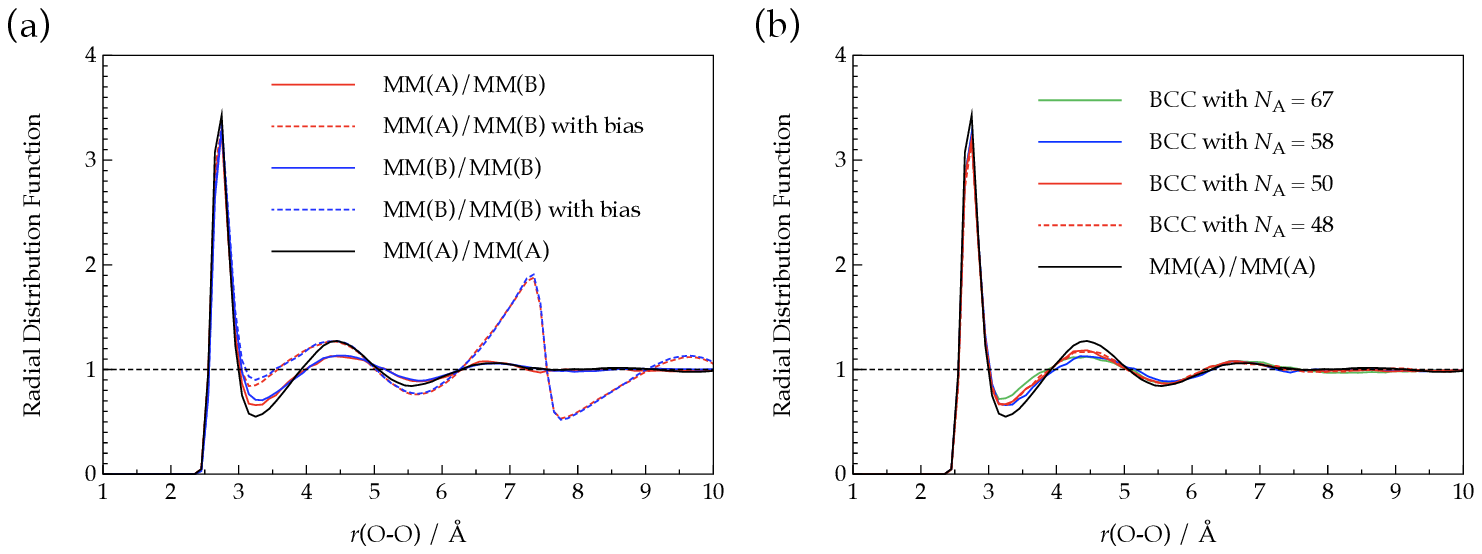}}            
\caption{\label{Rc=75A} (a) The radial distribution functions (RDFs) relevant to the BCC simulation are presented for the oxygen atoms of solvent($\rm{H}_2\rm{O}$) around the solute($\rm{H}_2\rm{O}$) oxygen. The radius of the shell $\Omega$ is set at $r_c=7.5$ \AA\; and $N_{\rm A}$ is set at 58. The notations in the legends are the same as those for Fig. \ref{Rc=35A}(a). \textcolor{black}{(b) The RDFs obtained by the BCC method are presented for the simulations with $N_{\rm A} = 48, 50, 58, {\rm and }\; 67$($r_c=7.5$ \AA). The RDF (\textquoteleft MM(A)/MM(A)\textquoteright\;) for the solution with pure solvent \textsf{A} is also shown as a reference.}}
\end{figure}

\indent We also applied the BCC method to a system with larger shell $\Omega$($r_c = 7.5$ \AA), for which $N_{\rm A}$ was set at $58$. The RDFs relevant to the BCC calculation is shown in Fig. \ref{Rc=75A}(a). A notable feature in the figure is again the artificial kinks in the RDFs (\textquoteleft ${\rm MM(A)}/{\rm MM(B)}$ with bias\textquoteright\; and \textquoteleft ${\rm MM(B)}/{\rm MM(B)}$ with bias\textquoteright) for the trajectories under the influence of the bias potential. The degrees of the rises and the falls before and after the O$-$O distance $r = 7.5\;$\AA\; are larger than those for the RDFs with a smaller shell $\Omega$. This may be attributed to the fact that the bias potential is exerted on the solvent with a large density almost comparable to the bulk density.  Anyway, the error is fully compensated by the correction of Eq. (\ref{eq:corr}) as exhibited in the graph of \textquoteleft ${\rm MM(A)}/{\rm MM(B)}$\textquoteright. However, the corrected RDF is rather closer to that for the pure SPC/E solvent (\textquoteleft ${\rm MM(B)}/{\rm MM(B)}$\textquoteright) than that for pure solvent \textsf{A}(\textquoteleft ${\rm MM(A)}/{\rm MM(A)}$\textquoteright). This shows a clear contrast to the previous result obtained for a smaller shell. To clarify the effect of the choice of the number $N_{\rm A}$ on the RDF we also performed sets of BCC simulations for $N_{\rm A} = 50$ and $67$. The results were compared with the RDF for $N_{\rm A} = 58$ as well as that for pure solvent \textsf{A}. It is seen in the figure that the BCC simulation with $N_{\rm A} = 50$ offers the closest RDF to that for pure solvent \textsf{A} among the three RDFs \textcolor{black}{though the difference between the plots for $N_{\rm A} = 50$ and $58$ is quite modest}. \textcolor{black}{To quantify the deviations of each RDF from the reference RDF \textquoteleft ${\rm MM(A)}/{\rm MM(A)}$\textquoteright, we calculated the root-mean-square deviations (RMSDs) for the region of $3.05$ \AA \; $\le$ \textit{r}(O-O) $\le$ 5.55  \AA \; which fully covers the second peaks of the RDFs. Then, the RMSDs were obtained as $0.071$, $0.086$, and $0.11$, respectively, for $N_{\rm A} = 50, 58,$ and $67$. } It is, thus, observed that the RDF deviates from the reference (pure solvent \textsf{A}) with the increasing number $N_{\rm A}$. As described in Computational Details the number $N_{\rm A} = 58$ for the construction of Fig. \ref{Rc=75A}(a) had been determined from a preliminary simulation for the solution with pure solvent \textsf{B}. More explicitly, the average number $N_{\rm av}$ of the water molecules within the shell was evaluated through the simulation, and then, the number $N_{\rm av}$ was taken as $N_{\rm A}$. We note, however, that the instantaneous number $N_{\rm in}$ of the molecules inside the shell substantially fluctuates during the simulation. Actually, the minimum and the maximum numbers of water molecules in the shell were $48$ and $69$ in our simulation, respectively. Therefore, confining  $N_{\rm av}$ solvent molecules constantly within the shell throughout the BCC simulation will give rise to some artifacts.  It is, thus, suggested that the number $N_{\rm A}$ for a given size of the shell should be chosen so that it allows the fluctuation of the number of the water molecules contained in the shell during a simulation. In other words, $N_{\rm A}$ should be appropriately smaller than $N_{\rm av}$ considering the deviation of  $N_{\rm in}$. The choice of $N_{\rm A} = 50$ is, thus, found to be suitable for the wall position of $r_c = 7.5$ \AA. \textcolor{black}{The calculation for $N_{\rm A} = 48$ was also performed and the result is plotted in Fig. \ref{Rc=75A}(b), which shows little difference with the plot for $N_{\rm A} = 50$ as expected.} We speculate that the naive choice of $N_{\rm A}= N_{\rm av}$ for the smaller shell (i.e. $r_c = 3.5$\; \AA) gave a rather successful result since $N_{\rm in}$ does not fluctuate largely. Anyway, the proper number $N_{\rm A}$ for a shell $\Omega$ can be readily determined through a preliminary calculation. \\
\indent \textcolor{black}{In closing this subsection we make a brief remark on the correction term of Eq. (\ref{eq:corr}). As demonstrated in the above simulations, the correction works adequately when the potential of \textsf{A} is reasonably close to that of \textsf{B}. However, it is possible that a flaw will emerge in the RDF when the force fields for \textsf{A} and \textsf{B} are largely different from each other. \textcolor{black}{Actually, the artifact of the bias potential remains in principle when solvent \textsf{A} is not identical to \textsf{B} in the BCC approach as in the other constraint approaches.} In the following subsection we also test the BCC correction by applying it to an actual QM/MM system involving an ionic molecule as a solute of interest. }

\subsection{QM/MM-BCC simulation for hydronium ion}
\begin{figure}[h]
\scalebox{0.43}[0.43] {\includegraphics[trim=0 0 0 0,clip]{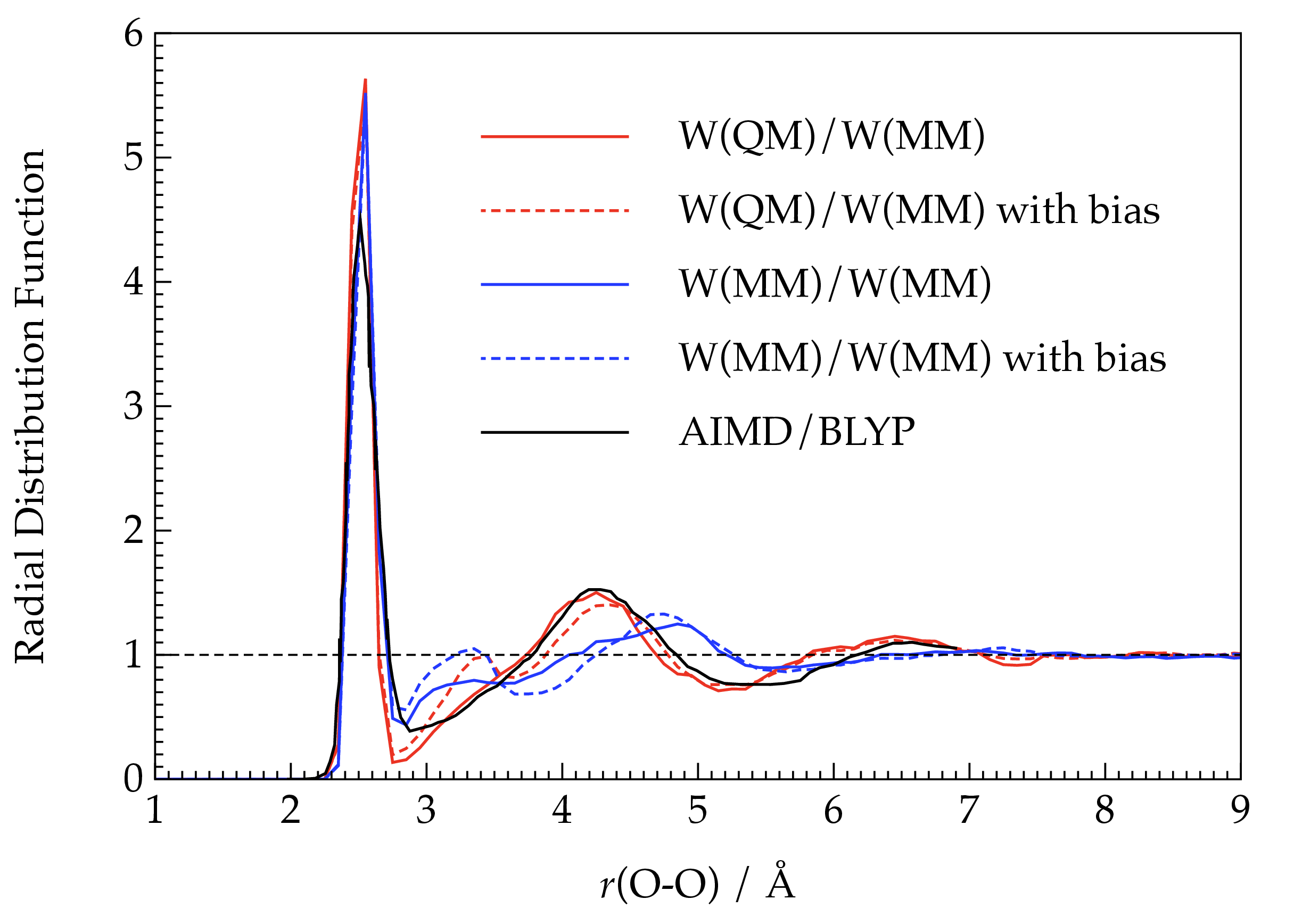}}            
\caption{\label{H3O+_BCC_rev}  The radial distribution functions (RDFs) relevant to the QM/MM-BCC simulation are presented for the oxygen atoms of water solvent around the oxygen of the solute (hydronium ion). The notations W(QM) and W(MM) stand for the quantum mechanical (QM) water (W) molecules and molecular mechanical (MM) water molecules, respectively. \textquoteleft W(QM)/W(MM)\textquoteright\; refers to the RDF obtained by the QM/MM-BCC simulation. The result given by an \textit{ab initio} molecular dynamics (AIMD) simulation is from the work of Tse, Knight, and Voth.\cite{Tse2015} }
\end{figure}

\begin{figure}[h]
\scalebox{0.43}[0.43] {\includegraphics[trim=0 50 0 0,clip]{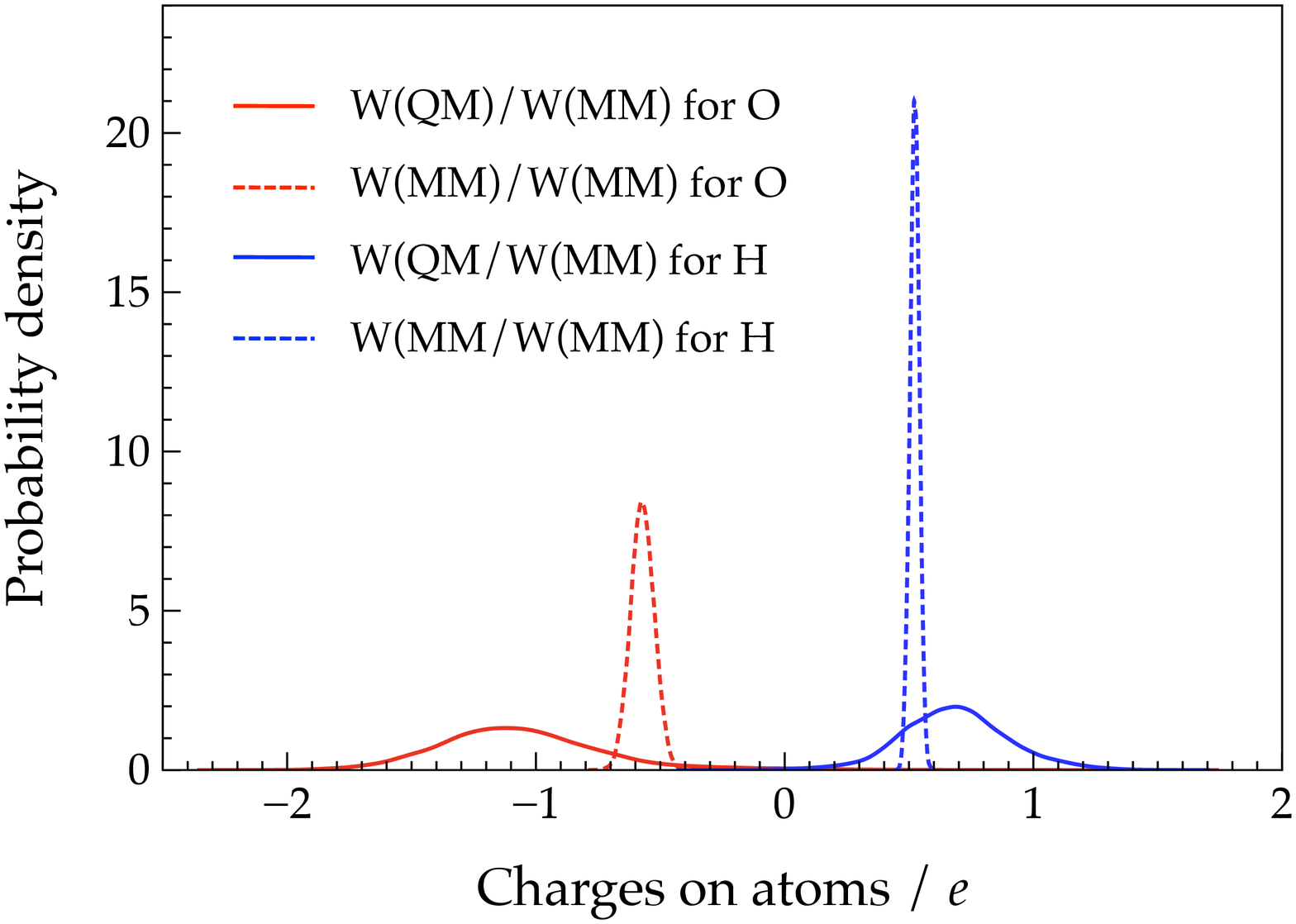}}            
\caption{\label{Charge}  Probability densities of the ESP(electrostatic potential) charges on the oxygen (O) and hydrogen (H) atoms on a hydronium ion in the trajectories of \textquoteleft W(QM)/W(MM) with bias\textquoteright\; and \textquoteleft W(MM)/W(MM) with bias\textquoteright. Charge value is given in the unit of the elementary charge $e$. \textcolor{black}{The ESP charges were determined so that the point charges on the atomic sites reproduce the actual electrostatic field formed by the instantaneous electron density of the QM objects through least-square fittings.} The sample points for the least-square fittings were randomly taken in the region outside the spheres with van der Waals radius of SPC/E centered at oxygen atoms.  }
\end{figure}

We combine the QM/MM simulation with the BCC approach (QM/MM-BCC) to study the hydration of a hydronium ion. The ${\rm H}_3{\rm O}^+$ and 4 water molecules confined within a shell $\Omega$ ($r_c = 3.5$\;\AA) were described by the KS-DFT method. The RDFs for the solvent oxygen around the hydronium oxygen are shown in Fig. \ref{H3O+_BCC_rev} where the result given by the \textit{ab initio} molecular dynamics (AIMD) simulation\cite{Tse2015} at the same thermodynamics conditions ($\rho = 1.0$ g/$\rm{cm}^3$, $T = 300$ K) is also provided as a reference. We note that the AIMD simulation in the graph was also yielded with the same exchange-correlation functional (BLYP) as our DFT calculation in the QM region. In Fig. \ref{H3O+_BCC_rev} the legend \textquoteleft W(QM)/W(MM) with bias\textquoteright, for instance, refers to a constrained QM/MM simulation where the 4 solvent water molecules are described with KS-DFT and the rest of the solvent are represented by the classical force field. Similarly, \textquoteleft W(MM)/W(MM)\textquoteright\; stands for the ordinary QM/MM simulation where all the solvent molecules are treated as MM objects.  It is seen in the figure that the RDF constructed by QM/MM-BCC (\textquoteleft W(QM)/W(MM)\textquoteright) shows good agreement with the AIMD result particularly in the region around the second peak. \textcolor{black}{The coordination number $N_c$ was estimated as $3.4$ for the RDF of \textquoteleft W(QM)/W(MM)\textquoteright\; by accumulating the population up to 3.0 \AA . And the RDF for AIMD simulation provided $N_c = 3.5$, showing a rather good agreement with QM/MM-BCC. On the other hand, we found the conventional QM/MM approach yielded a slightly larger value ($N_c = 3.7$), which can be attributed to the larger population at the minimum around $r$(O-O) = 3.0 \AA. } It is also worthy of note that the height and the position of the second peak in QM/MM-BCC as well as in AIMD are rather different from those in the ordinary QM/MM (\textquoteleft W(MM)/W(MM)\textquoteright). The origin of the difference would be attributed to the polarization of the QM solute and the surrounding QM water molecules in the QM/MM-BCC and AIMD simulations. To substantiate this speculation the probabilities of the ESP(electrostatic potential) charges on the oxygen and hydrogen atoms on the hydronium ion in the trajectories of \textquoteleft W(QM)/W(MM) with bias\textquoteright\; and \textquoteleft W(MM)/W(MM) with bias\textquoteright\; are plotted in Fig. \ref{Charge}. It can be readily recognized in the figure that the electrons on the solute surrounded by the QM water molecules are more polarized and fluctuate more significantly as compared with the solute embedded in the pure classical solvent suggesting the importance of the charge-transfer type polarization between solute and solvent in the hydration of an ion.  
 As a consequence the average of the charges on the oxygen atoms on the QM water molecules was enhanced to $-1.0357$ in the unit of the elementary charge. Thus, the QM water molecules in the first solvation shell are more polarized than the SPC/E water, \textcolor{black}{though it is not straightforward to compare the QM charges with empirical MM partial charges.} Therefore, it is possible that the water molecules relevant to the formation of the second peak will be attracted more strongly to the first solvation shell, which leads the decrease in the O-O distance and the enhancement of the distribution at the second peak. \textcolor{black}{Of course, the water molecules in the second solvation shell are not treated quantum mechanically in our QM/MM-BCC simulation in contrast to the AIMD simulation. However, the SPC/E model of water will work adequately outside the first solvation shell since the effect of the ionic solute might be rather weakened. }It should also be noted that the first peak in the RDF of QM/MM-BCC is narrower and its height is more emphasized than those of AIMD.  We speculate that the distributions will be more broadened in the AIMD simulations since the geometries of the individual solvent molecules as well as the solute were being flexible. It was, thus, demonstrated that the QM/MM-BCC method can reasonably realize the solvation structure around an ion in comparable accuracy with the result given by a first-principle simulation. 

\begin{figure}[h]
\scalebox{0.43}[0.43] {\includegraphics[trim=0 0 0 0,clip]{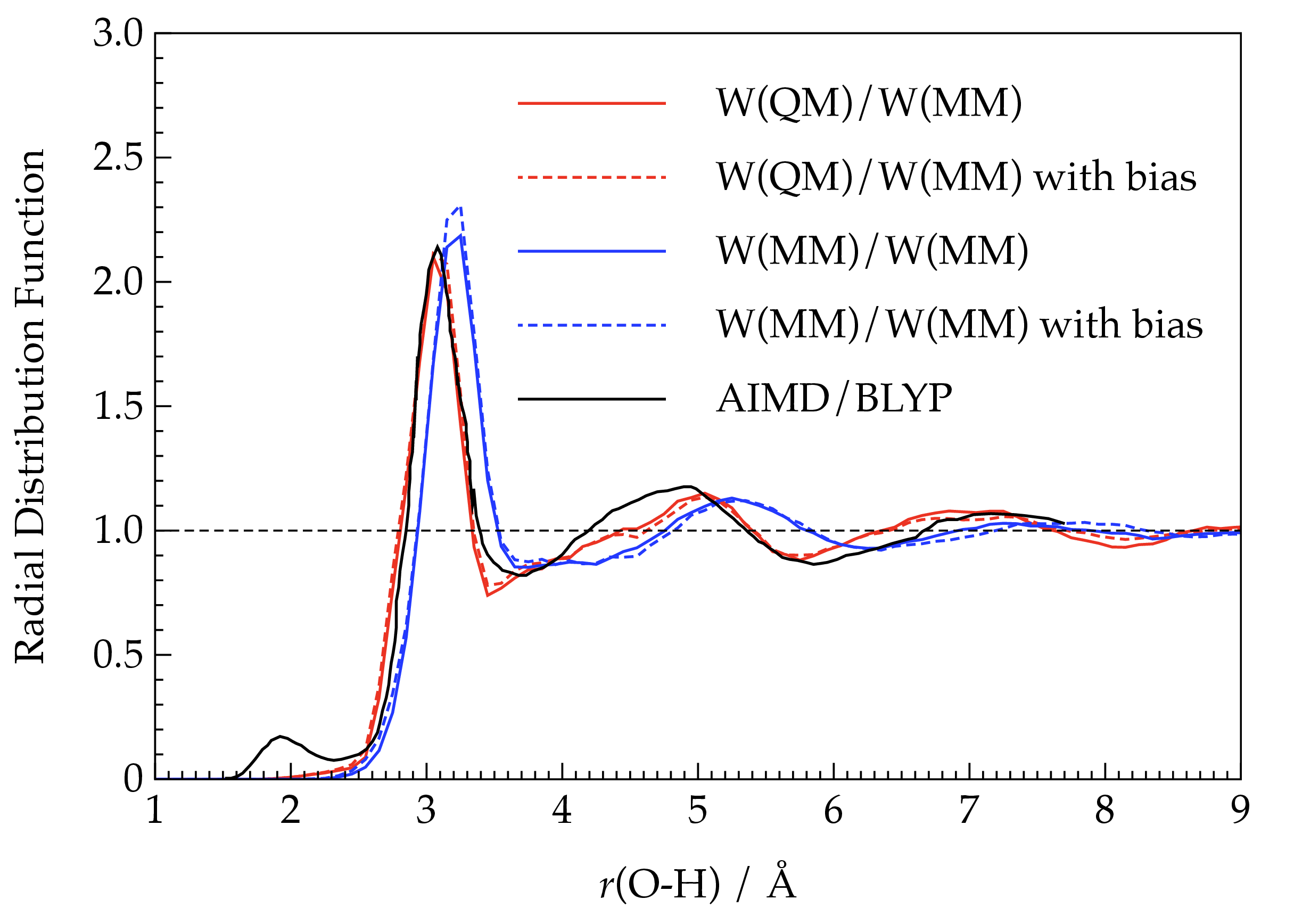}}            
\caption{\label{H3O+_BCC_OH}  The radial distribution functions (RDFs) relevant to the QM/MM-BCC simulation are presented for the hydrogen atoms of water solvent around the oxygen of the solute (hydronium ion). The notations W(QM) and W(MM) stand for the quantum mechanical (QM) water (W) molecules and molecular mechanical (MM) water molecules, respectively. \textquoteleft W(QM)/W(MM)\textquoteright\; refers to the RDF obtained by the QM/MM-BCC simulation. The result given by an \textit{ab initio} molecular dynamics (AIMD) simulation is from the work of Tse, Knight, and Voth.\cite{Tse2015} }
\end{figure}

\indent \textcolor{black}{Lastly we discuss the RDFs for the hydrogen atoms of solvent around the oxygen of the hydronium ion. The plots are presented in Fig. \ref{H3O+_BCC_OH} where the RDF given by the AIMD simulation\cite{Tse2015} is also shown to make comparisons. We found that the RDF of QM/MM-BCC (\textquoteleft W(QM)/W(MM)\textquoteright\; ) shows a good agreement with that of AIMD in the first prominent peak. The origin of the small peak around $r$(OH) = 2.0 \AA\; in AIMD was fully discussed in Ref. \onlinecite{Tse2015} and attributed to the water molecules donating a weak H-bond to the oxygen of the hydronium ion. Importantly, it was observed that the donation of the hydrogen atom is strongly coupled with the dynamics of the proton transfers occurring in the AIMD simulation that are, of course, not allowed in the present QM/MM-BCC simulations. We also see that the second peak of QM/MM-BCC reasonably agrees with that of AIMD.  It is recognized in the figure that the effect of the bias potential on the O$-$H RDF is marginal in contrast to the O$-$O RDF. This would be due to the fact that the bias potential is exerted only on the oxygen atoms and not directly on the hydrogen atoms. It is also worth noting that the position of the second peak of the conventional QM/MM(\textquoteleft W(MM)/W(MM)\textquoteright\ )  is somewhat different from that of QM/MM-BCC . The source of the discrepancy can also be explained in the same way as in the O$-$O RDF. That is, the solvent molecules in the second shell in QM/MM-BCC will be attracted more strongly to the solute due to the enhanced polarization of the first shell.  }  

\section{CONCLUSION}
We developed in this article a simple and efficient method categorized in the constrained QM/MM simulation. The point of our method called BCC (boundary constraint with correction) is to compensate the error due to the bias potential by adding a correction term obtained through a set of separate QM/MM simulations. The BCC approach fulfills the desirable conditions that the energy and forces are continuous and the energy and the momentums are conserved. Furthermore, the method is designed so that the effects of the applied bias potential for constraint completely disappears when the QM solvent subjected to the bias potential is identical to the MM solvent. In the BCC method various types of constraint potential will be used and various forms of the shell $\Omega$ can be adopted according to the shape of the solute molecule of interest.  

\indent As benchmark tests we applied the BCC method to two kinds of water solutions. First we considered a solute water molecule embedded in an aqueous solution represented with a fully classical force field.  We, then, computed the O-O RDFs by means of the BCC procedures for two different sizes of $\Omega$ ($r_c = 3.5$\; and $7.5$\AA). It was demonstrated that the RDFs given by the BCC method were in good agreements with those for the references. It should be kept in mind, however, that the choice of the number $N_{\rm A}$ of the solvent molecules for a given size of the shell will somewhat affect the BCC results. Our recommendation is to take the appropriately smaller value than the average number $N_{\rm av}$ of the solvent molecules inside the shell. More explicitly, $N_{\rm A}$ should be determined so that it allows the deviation of the instantaneous number $N_{\rm in}$ in the shell from $N_{\rm av}$. The number $N_{\rm av}$ and the deviation can be readily obtained through a preliminary simulation. Further, we combined the QM/MM simulation with the BCC method (QM/MM-BCC) and applied it to the hydration of a hydronium ion. The O-O RDF for the solvent around the solute was computed using our method, which shows a fairy good agreement with a reference RDF given by an AIMD simulation.  

\indent Thus, it was shown that the BCC approach is simple yet effective and robust in computing a statistical property of a QM/MM system subjected to a constrained force. Our next issue as an extension of QM/MM-BCC is to compute the solvation free energy of an anionic QM solute in an aqueous solution. The formulation of the method and the test calculations are now proceeding. The results will be reported in forthcoming issues.

\begin{acknowledgments}
This work was supported by the Grant-in-Aid for Scientific Research on Innovative Areas (No. 23118701) from the Ministry of Education, Culture, Sports, Science, and Technology (MEXT), by the Grant-in-Aid for Challenging Exploratory Research (No. 25620004) and the Grant-in-Aid for Scientific Research(C) (No. 17K05138) from the Japan Society for the Promotion of Science (JSPS). The calculations were performed partly using computational resources of the HPCI systems provided by SX-ACE in Osaka University and in Tohoku University, and Cray XC30 at Kyoto University through the HPCI System Research Project (Project IDs: hp150131, hp160007, hp160013, and hp170046).
\end{acknowledgments}


\bibliographystyle{apsrev4-1}

\end{document}